\documentclass[preprint]{elsarticle}
\usepackage{color}
\usepackage{array}
\usepackage{bm}
\usepackage{amsmath} 
\usepackage{amsfonts} 
\usepackage{graphicx} 
\usepackage[utf8]{inputenc}
\usepackage{natbib}
\usepackage{amssymb}
\usepackage[normalem]{ulem}

\begin{document}

\title{The role of the number of degrees of freedom and chaos in  
       macroscopic irreversibility}

\author[uni,cnr]{L. Cerino}\ead{luca.cerino@roma1.infn.it}

\author[cnr]{F. Cecconi}
\ead{fabio.cecconi@roma1.infn.it}

\author[cnr]{M. Cencini}
\ead{massimo.cencini@roma1.infn.it}

\author[uni,cnr]{A. Vulpiani}
\ead{angelo.vulpiani@roma1.infn.it}
\address[uni]{Dipartimento di Fisica, Universit\`a Sapienza di Roma, p.le Aldo Moro 2, 00185 Roma, Italy}
\address[cnr]{CNR-ISC, Istituto dei Sistemi Complessi, Via dei Taurini 19, 00185 Roma, Italy}

\begin{abstract}
This article aims at revisiting, with the aid of simple and neat numerical
examples, some of the basic features of macroscopic irreversibility,
and, thus, of the mechanical foundation of the second principle of
thermodynamics as drawn by Boltzmann. 
Emphasis will be put on the fact that, in systems characterized by a very  
large number of degrees of freedom, irreversibility is already 
manifest at a single-trajectory level for the vast majority of the
far-from-equilibrium initial conditions -- a property often referred to as 
{\it typicality}.
We also discuss the importance of the
interaction among the microscopic constituents of the system and the
irrelevance of chaos to irreversibility, showing that the same
irreversible behaviours can be observed both in chaotic and
non-chaotic systems.
\end{abstract}
\maketitle

\section{Introduction}
Everyday experience demonstrates that many natural processes are
intrinsically irreversible at the \textit{macroscopic
level}. Think of a gas initially confined by a septum
in one half of a container, that spontaneously fills the whole available volume
as soon as the separator is removed.  Or, closer to daily experience,
consider the evolution of an ink drop into water \cite{LEBO93}. We
would be astounded and incredulous while observing the reverse
processes to occur spontaneously: a gas self-segregating in one half
of the container, or an ink drop emerging from a water-and-ink
mixture. In thermodynamics, the second principle amounts to a
formalization of this state of ``incredulity''.  From Newtonian (and
quantum) mechanics, we know that at the
\textit{microscopic level} the dynamics is reversible. How can we 
reconcile macroscopic irreversibility with microscopic reversibility
of the dynamics ruling the elementary constituents of macroscopic
bodies?

A solution to this riddle was proposed more than 140 years ago, when
Boltzmann laid down the foundation of statistical mechanics. At the
beginning, Boltzmann's ideas on macroscopic irreversibility elicited a
heated debate mainly due to the \textit{recurrence paradox},
formulated by Zermelo, and the \textit{reversibility paradox} by
Loschmidt (a detailed discussion on the historical and conceptual
aspects of the Boltzmann's theory can be found in
Refs. \cite{CERC06,CRV2014}). In spite of several rigorous
mathematical results \cite{CERC06,LANF75} supporting, with a clear
physical interpretation (at least for many scientists, including the
authors of this paper), the coherence of the scenario proposed by
Boltzmann, irreversibility still remains a somehow misinterpreted and
controversial issue, even among researchers in the field. The reader
may appreciate some of the opinions from the comments~\cite{BCFSDH94}
to a paper by Lebowitz on Boltzmann's ``time arrow'' ~\cite{LEBO93}.
 
According to Boltzmann, irreversibility is well defined only for
systems with a very large number of degrees of freedom. It
should be observed in the vast majority of the
individual realizations of a macroscopic system starting far from
equilibrium: ``vast majority'' is usually referred to
as \textit{typicality} in the literature \cite{Goldstein2012,Zanghi05}
(see next section for further details). Hence, there is no need to
repeat the experiment many times to understand that the free-gas
expansion or the spreading of an ink drop in the water are
irreversible processes, a single observation is enough.  Conversely,
for some authors irreversibility can only be properly
defined through the use of ensembles. Also, there is
not general agreement on the fact that irreversibility is an emergent
property when the number of degrees of freedom becomes
(sufficiently) large. For instance, Prigogine and his school claim
that {\it Irreversibility is either true on all levels or on none: it
cannot emerge as if out of nothing, on going from one level to
another~\cite{PS84}.}   For others, irreversibility results from
(microscopic) chaotic dynamics, or, it is a mere consequence of the
interaction with the external environment.

Likely due to this maze of different opinions, often there is a
persistence of confusing and conflicting ideas about macroscopic
irreversibility in spite of clear discussions
of the subject in the recent past \cite{LEBO93,BRIC95,Goldstein2001}.

This article aims at supplementing some aspects of Boltzmann's 
explanation of macroscopic
irreversibility which are often sources of misinterpretation with
simple and neat numerical examples.  In particular, the article
focuses on the behavior of macroscopic observables that can be
measured in laboratory experiments. Boltzmann's solution to the
reversibility and recurrence paradoxes will be mainly left aside 
as it already widely discussed in the literature (see, e.g., \cite{huang,KAC57}). Instead,
emphasis will be put on the fact that irreversibility is a property of
the single realization of a macroscopic system in which an important
role is played by the (even very weak) interaction among its
elementary constituents. In fact, although irreversible behavior can
be manifest also in systems of non-interacting units (see
Ref.~\cite{SWE08} for a pedagogical presentation of irreversibility in
the case of non-interacting gas free expansion), in the absence of
interactions single-particle and ensemble properties trivially
coincide leading to some ambiguity in the interpretation of
irreversibility. Finally, to stress the generality of the ideas,
simple models, which can be studied by standard simulation techniques,
will be considered. In particular, the comparison between chaotic and
non-chaotic systems will underline the irrelevance of chaos to
irreversibility.

The paper is organized as follows. In Section~\ref{sec:2} we 
briefly survey the conceptual aspects of Boltzmann's approach, and
discuss the entropy, the use of ensembles and the role of chaos.
Moreover we present some explicit calculations performed in a
Markovian model introduced by P. and T. Ehrenfest
\cite{EE911}. In Sect.~\ref{sec:3}, we consider numerical examples of 
macroscopic irreversible behaviors by studying deterministic systems
involving particles which collide with a moving wall.  In
section~\ref{sec:4}, we discuss a numerical experiment conceptually
alike to the irreversible mixing of an ink drop into water.
Section~\ref{sec:5} summarizes the main aspects of our understanding
of macroscopic irreversibility on the light of Boltzmann's ideas.

\section{Basic facts on macroscopic irreversibility\label{sec:2}}

We start recalling some basic notions. In Classical Mechanics
(quantum systems will not be treated here) a macroscopic body, 
is fully described once we specify its microstate
$\bm X(t)\equiv(\bm x_1(t);\ldots; \bm x_N(t))\equiv 
(\bm q_1(t),\bm p_1(t);\ldots; \bm
q_N(t), \bm p_N(t))$ characterized by the position $\bm q_i$ and the 
momentum $\bm p_i$ of its $N$ elementary constituents, say particles.
The whole set of admissible microscopic configurations, $\{\bm X\}$, 
defines the phase space, or
$\Gamma$-space.  The evolution of a macroscopic system from
an initial state $\bm X(0)$ at time $0$ up to a specified time $T>0$,
$\{\bm X(t)\}_{t=0}^{T}$, constitutes a ``forward''
trajectory. The time ``reversed'' trajectory is
obtained by applying the time reversal transformation $\mathcal{R}$,
i.e. considering as initial state the one with particles at the
positions reached at time $T$ but with reversed velocities, i.e. $\bm
X^{\mathcal{R}}(0)=\mathcal{R} (\bm X (T))\equiv (\bm q_1(T),-\bm
p_1(T);\ldots; \bm q_N(T), -\bm p_N(T))$.  When the system is evolved
from $\bm X^{\mathcal{R}}(0)$, thanks to the invariance of Newton's
equations under time reversal, it traces back the forward trajectory
(with reversed velocities) as if the evolution movie were played
backwards, i.e. given $0\le t\le T$, $\bm
X^{\mathcal{R}}(t)=\mathcal{R} (\bm X(T-t)$).

From Thermodynamics, we know that the macrostate of a 
large system ($N\gg 1$) is specified by a small number of macroscopic
observables, $M_\alpha(t)=M_\alpha(\bm X(t))$ with
$\alpha=1,\ldots,k\ll N$. 
The observables $M_\alpha$ to be qualified as ``macroscopic'' 
must depend on a large number of the system degrees of freedom. 
In general, we have that many microscopic
configurations correspond to the same value of the observables, in
other terms the relation between micro and macro state is many to
one. Some examples are the energy of a subsystem composed of many
particles, the number density in specific (not too small) regions, or
the number of particles with velocity in a given interval.  At
equilibrium the macroscopic observables assume specific values
$M_\alpha^{eq}\equiv \langle M_\alpha \rangle_{eq}$, where
$\langle \cdot \rangle_{eq}$ denotes the equilibrium average with
respect to, e.g., the microcanonical distribution (in principle, other
ensembles can be used, we use here the microcanonical one as it is the
appropriate one for the numerical examples discussed in the next
sections). 
We can define a state to be far from equilibrium when the observables 
deviate from their equilibrium values well beyond the equilibrium fluctuations, in other
terms when $||M_\alpha-M_\alpha^{eq}||\gg
\sigma^{eq}_M\equiv \sqrt{\langle M_\alpha^2\rangle_{eq} -
(M^{eq}_\alpha)^2}$. Conversely, whenever $||M_\alpha-M_\alpha^{eq}||\approx
\sigma^{eq}_M$ we speak of close-to equilibrium states.

Macroscopic irreversibility refers to the fact that 
when starting from far-from equilibrium states, the (macroscopic)
system evolves toward equilibrium, i.e. at times long enough we have
that $M_\alpha(t) \to M_\alpha^{eq}$, while we never observe the
opposite, i.e. that starting close to equilibrium the system 
approaches (spontaneously) a far from equilibrium state, in spite of the
fact that such reversed trajectories would be perfectly compatible
with the microscopic dynamics.\footnote{Obviously,
weakly interacting particles, in an empty infinite space, can
spontaneously leave the region where they were initially released and
never return there \cite{SWE08}. This form of irreversibility is quite
trivial, so we shall only consider systems evolving in a bounded
region of $\Gamma$.} 

Boltzmann explained the asymmetry in the time
evolution of macroscopic systems in term of a probabilistic
reasoning. He realized that the number of microscopic configurations
corresponding to the equilibrium state, i.e. $\bm X$ such that
$M_\alpha(\bm X) \approx M_\alpha^{eq}$ is, when the number of degrees
of freedom $N$ is very large, astronomically (i.e. exponentially in
$N$)\footnote{Since in macroscopic bodies $N$ is order
the Avogadro number, $N_A\approx 10^{23}$ we are speaking here of hard
to imagine larger numbers when the exponential is taken.} larger than
those corresponding to non-equilibrium states. Somehow
``intuitively'' it is overwhelmingly ``more probable'' to see a system
evolving from a very ``non-typical'' state, i.e. which can be obtained
with (relatively to equilibrium) a negligible number of microscopic
configurations, towards an equilibrium state, which represents a huge
number of microscopic states, than to see the opposite. This
``intuitive''\footnote{Intuitive only a posteriori and
in a very subtle way indeed.}  notion of ``more probable'' can be
formalized in terms of the Boltzmann's entropy of a given macrostate,
which is the log of the number of microstates corresponding to that
macrostate, one of the greatest contribution of Boltzmann was to
identify such entropy with the thermodynamic entropy when in
equilibrium.  These entropic aspects have been (beautifully and
thoroughly) discussed in other
articles~\cite{LEBO93,BRIC95,Goldstein2001}, to which we refer to.

In the case of very dilute (monoatomic) gases, Boltzmann was even able to do more, with his celebrated $H$-theorem, by demonstrating the irreversible dynamics of the one-particle empirical distribution function\footnote{Here, we define it through Dirac-deltas from a mathematical point of view we should always think to some regularization via, e.g., some coarse-graining.} $f_1(\bm
x,t)=\frac{1}{N}\sum_{i=1}^{N} \delta(\bm x-\bm x_i(t))$\,, where
${\bm x}=(\bm q,\bm p)$ denotes the position and momenta of a single
particle, i.e. the so-called $\mu$-space. The interesting aspects
about the empirical distribution are that $f_1$ is a well defined
macroscopic observable and can be, in principle, measured in a single
system, e.g. in numerical simulation. In an appropriate asymptotics
(the so-called Boltzmann-Grad limit, see Ref.\cite{CERC06} for
details) the evolution of $f_1$ is well described by a deterministic
equation -- the Boltzmann's equation. This equation, via
standard derivations (see, e.g., Ref. \cite{huang, KAC57}) predicts
an asymmetry in the evolution of the quantity
\begin{equation}
H(t)=k_B \int f_1({\bm x}, t) \ln f_1({\bm x},t) d{\bm x} \,.
\label{eq:sb}
\end{equation}
  In other terms, the $H$-theorem states that if the system is truly
macroscopic, i.e. $N$ is huge (which allows us to consider a single
system and to describe it at a macroscopic level by using the
empirical distribution), if its initial state is far from equilibrium,
the function $H(t)$ cannot increase (but for small
fluctuations) \cite{huang,CERC06,LEBO93}. In particular, it is maximal
at equilibrium, which for a dilute gas implies uniform distribution in
the spatial coordinate and Maxwell-Boltzmann distribution for the
velocities, where it is nothing but minus the Boltzmann's entropy
$S_B(t)=-H(t)$ and thus the thermodynamic entropy.

The main criticisms to the first formulation of the $H$-theorem boil
down to the well known reversibility and recurrence paradoxes. The
former, formulated by Loschmidt, states that the invariance under time
reversal of Newton's mechanics implies that time-reversed
trajectories have nothing exceptional (from a microscopic point of
view) with respect to forward ones, so that such reversed trajectories
can be used to ``invert'' the theorem and thus to show
that $H(t)$ must increase, i.e. the entropy must
decrease. The criticism by Zermelo was based on Poincar\'e recurrence
theorem: the state of a mechanical system, evolving in a bounded phase
space region, will return infinitely close to the initial state, so
that there will be a time at which $H(t)$ will come
back to the original value, again contradicting the theorem.

The Boltzmann's solution to these paradoxes has been discussed in many
texts and manuals (see e.g. \cite{huang,KAC57}), and thus will not be
discussed in details here. We simply recall that,
given the macroscopic nature of the system the Poincar\'e time can
much larger than the age of the universe in a true macroscopic body
and that, as mentioned before, the number of microstate corresponding
to equilibrium is astronomically large with respect to those far from
equilibrium, justifying the typicality of macroscopic
irreversibility.

In the sequel we shall focus, within the framework of specific
examples, on the fact that macroscopic irreversibility is well defined
in a single realization (i.e. no need to average over
the initial probability density), which is again a manifestation of
the aforementioned typicality. Before entering the specific examples,
it is useful to briefly recall some ideas about the role of ensembles
and chaos on the notion of irreversibility, as their relevance to the
latter might be subject of a certain confusion.

\subsection{Ensembles, chaos and entropy}\label{sec:2a}

Although the importance of probabilistic methods in statistical mechanics
cannot be underestimated, it is necessary to answer the following
question: what is the physical link between the probabilistic
computations (i.e. the averages over an ensemble) and the actual
results obtained in laboratory experiments
which, \textit{a fortiori}, are conducted on a single realization (or
sample) of the system under investigation?

The answer of Boltzmann is well captured by the notion
of \textit{typicality} \cite{Goldstein2012}, i.e. the fact that the
outcome of an experiment on a macroscopic system takes a specific
(typical) value overwhelmingly often. In statistical mechanics
typicality holds in the thermodynamic limit (and thus for $N\gg
1$). It is in such an asymptotics that the ratio between the set of
typical (equilibrium) states and non-typical ones goes to zero
extremely rapidly (i.e. exponentially in $N$), thus it is only when
$N$ is large that the probability to see the irreversible dynamics of
initially far-from equilibrium macrostates towards equilibrium ones
becomes (at any practical level) one. The concept of typicality is not
only at the basis of the second law, but (possibly at a more
fundamental level) in the very possibility to have reproducibility of
results in experiments (on macroscopic objects) or the possibility to
have macroscopic laws~\cite{BRIC95}.  Consider a system with $N$
particles, and a given macroscopic observable $M_N({\bm X})$.  Let us
assume an initial well behaving\footnote{From a
mathematical point of view this means that it has to be absolutely
continuous with respect to the Lebesgue measure.} phase-space density
$\rho({\bm X},0)$ prescribing a given macroscopic state. From a
physical point of view we can assume, e.g, $\rho({\bm X},0)=0$ if
$M_N({\bm X})\notin [M_0:M_0+\Delta M]$, for some $M_0$ (usually
chosen far from equilibrium) with $\Delta M/M_0 \ll 1$, that is we
consider that one or more (macroscopic) constraints on the dynamics
are imposed.  Then we consider the ensemble of the microstates
compatible with that constraint. Common examples are, e.g., a gas at
equilibrium confined in a portion of the container by some separator
(see next Section for some numerical examples).  At time $t=0$ such
constraints are released and we monitor the evolution of the system by
looking at the macroscopic observable $M_N({\bm X}(t))$: we denote
with $\langle M_N(t)\rangle$ the average over all the possible initial
conditions weighted by $\rho({\bm X},0)$.  If $N\gg 1$ and the initial
state is far from equilibrium $||M_0-M^{eq}||\gg \sigma^{eq}_M$,
according to the ``Boltzmann's interpretation'' of irreversibility, the
time evolution of $M_N(t)$ must be typical i.e. apart from a set of
vanishing measure (with respect to $\rho({\bm X}(0),0)$), most of the
initial conditions originate trajectories over which the value of
$M_N({\bm X}(t))$ is very close to its average $\langle M_N(t)\rangle$ at every 
time $t$.\footnote{Such property does not hold for all the
observables in all situations, for instance one has to exclude
situations in which the macroscopic dynamics is unstable. In this case
the transient to equilibrium may vary from realization to realization
though the final equilibrium state will be reached by almost all the
realizations.}  In other terms, if $N$ is large, behaviors very
different from the average one (e.g. a ink drop {\it not} spreading in
water) {\it never} occur:
\begin{equation}
\text{Prob}\{\; M_N(t)\approx \langle M_N(t)\rangle\;\} 
\approx 1 \qquad \text{when}\qquad N\gg 1\,.
\label{eq:typ1}
\end{equation} 
The rigorous proof of the above conjecture is very difficult and, of
course, it is usually required to put some restrictions.  It is
remarkable that, as we will see in the next subsection, it is possible
to show the validity of this property in some stochastic systems.

The use of probability distribution to introduce the idea of
typicality, as from the discussion above, should not convince the reader
that irreversibility is a probabilistic notion. 
In particular, one should be careful to avoid the confusion between  
irreversibility and relaxation of the phase space probability
distribution.  If a dynamical system exhibits ``good chaotic
properties'', more precisely, it is \textit{mixing}, a generic
probability density distribution of initial conditions, the {\em
ensemble}, $\rho({\bm X},0)$, relaxes (in a suitable technical sense)
to the invariant distribution for large times $t$
\begin{equation}
\label{eq:mixing}
\rho({\bm X},t) \to \rho^{inv}({\bm X})\,\,.
\end{equation}
It is worth remarking that in systems satisfying Liouville theorem,
the relaxation to the invariant distribution must be interpreted in a
proper mathematical sense: for every $\epsilon>0$ and for every ${\bm
X}$, one has
\begin{equation}\label{eq:mixing-smooth}
\int_{|{\bm X-\bm Y}|<\epsilon}\rho({\bm Y},t)d{\bm Y} \rightarrow \int_{|{\bm X-\bm Y}|<\epsilon} \rho^{inv}({\bm Y})d{\bm Y}\,.
\end{equation}

We want to make clear here that the property (\ref{eq:mixing}) 
or, equivalently, (\ref{eq:mixing-smooth}) is a form of irreversibility 
completely unrelated to the second law of thermodynamics. In fact, it does not
require large systems as it can be observed even in dynamical systems
with few degrees of freedom (see also the discussion on
Sect.~\ref{sec:4}), for which no meaningful set of macroscopic
observables can be defined.

It is worth reporting that some authors have a different opinion.  For
example, in his comment to Lebowitz paper, \cite{LEBO93}
Driebe \cite{BCFSDH94} states that irreversible processes can be
observed in systems with few degrees of freedom, such as the baker
transformation or other reversible, low-dimensional chaotic systems.
However, one must appreciate that, in such low-dimensional 
chaotic systems, irreversibility due to the mixing property 
is observed only by considering ensembles of initial conditions, 
while single realizations do not show a preferential direction of time. 
This occurs also 
in macroscopic systems when we monitor 
the evolution of an observable that is not macroscopic, 
e.g. a single molecule property either in the gas or in the ink drop.
In that case, nothing astounding happens by looking at the forward or 
reversed trajectory, as we cannot decide the direction of the process.
For a critical discussion of the role of chaos in
irreversibility see Ref.~ \cite{BRIC95}.

A trivial consequence of interpreting Eq.~(\ref{eq:mixing}) and
(\ref{eq:mixing-smooth}) as a form of irreversibility is that systems
of $N\gg1$ non interacting particles, with a chaotic behavior, would
exhibit irreversibility, also in the thermodynamic
sense \cite{BIND99}. However it is clear that this cannot be the
case: in fact, some sort of (even weak) interaction among the
particles is necessary to observe genuine thermodynamic
behaviors and thus irreversibility.  This can be easily understood
considering a system with $N\gg1$ independent particles in a box:
suppose that the initial velocities of the particles labeled by
$i=1,\ldots,N/2$ are extracted from a Maxwell-Boltzmann distribution
at the temperature $T_1$, and that the others, $i=N/2+1,\ldots,N$, are
extracted from the same distribution, but at a different temperature
$T_2$. In the absence of interaction, the absolute value of the
momentum of each particle $|{\bm p}_i|$ does not change and, as a
consequence, the time evolution of some macroscopic observables
(e.g. $M(\bm X)=\frac{1}{N} \sum_i |{\bm p}_i|^4$) does not tend to
the microcanonical equilibrium value.

Such an elementary remark underscores that some degree of interaction
among particles constitutes an unavoidable ingredient for a correct
thermodynamic behavior.

In discussing irreversibility, some authors define the entropy using
the probability distribution function (PDF) in the $\Gamma$-space,
$\rho(\bm X,t)$.
This way one obtains the so-called Gibbs entropy
\begin{equation}
S_G(t) = -k_B \int \rho({\bm X}, t) \ln \rho({\bm X},t) d{\bm X} \,.
\end{equation}
However, $S_G$ can only be
defined over an ensemble, otherwise $\rho$ is meaningless. As a
consequence, $S_G(t)$ is accessible only in numerical experiments with
systems composed by few degrees of freedom. But, more crucially, it is
unclear how to relate $S_G$ to irreversibility because Liouville
theorem implies that $S_G(t)$ must stay constant over time!

In order to observe an increase over time for $S_G$-like quantities,
many authors introduce a coarse-graining of the $\Gamma$-space,
amounting to consider a partition of the phase space in cells of size
$\epsilon$ and to define the probability $P_j(t,\epsilon)$ that the
state $\bm X$ visits the $j$-th cell at time $t$. In this way we
obtain the coarse-grained Gibbs entropy
\begin{equation}
S_G^{(cg)}(t,\epsilon) = -k_B \sum_j P_j (t,\epsilon) \ln
P_j(t,\epsilon) \,\,.
\label{eq:sgcg}
\end{equation}
Now for $\epsilon>0$, $S_G^{(cg)}$ turns to be an increasing function
of time. However, it can be numerically shown that, for $\epsilon >0$,
the quantity $S_G^{(cg)}$ remains constant up to a crossover time
$t_* \sim \ln (1/\epsilon)$, after which it starts
increasing. Clearly, this $\epsilon$-dependence indicates that the
growth is a mere artifact of the coarse-graining and it is unrelated
to irreversibility, though it can be of some interest in the study of
dynamical systems \cite{CFLV_book}.

\subsection{Typicality and irreversibility in the  Ehrenfest model
\label{sec:eremodel}}

Let us now briefly discuss the meaning of typicality in a simple
stochastic example, where explicit computations can be performed. This
simple Markov chain was introduced by P. and T. Ehrenfest \cite{EE911}
to illustrate  some aspects of Boltzmann's ideas on irreversibility.  
According to Kac \cite{KAC57} {\it this Markov
chain is probably one of the most instructive models in the whole of
Physics and, although merely an example of a finite Markov chain, it
is of considerable independent interest.}

Consider $N$ particles, each of which can be either in one box ($A$)
or in another ($B$). The state of the Markov chain at time $t$ is
identified by the number, $n_t$, of particles in $A$ and the evolution
of the state is stochastic. The transition probabilities for the state
$n_t=n$ to become $n_{t+1}=n \pm 1$ are given by
\begin{equation}
P_{n\to n-1} =\frac{n}{N} \quad \mbox{and} \quad P_{n\to n+1}
=1-\frac{n}{N}
\,,\label{eq:trans}
\end{equation}
respectively.

We can now re-interpret the model in the language of statistical
mechanics. The state of the Markov chain $n_t=n$, at time $t$, can be
seen as the ``macroscopic'' state ($M$) of the system, the
corresponding ``microscopic'' configuration is defined by the
(labeled) particles which are effectively in box $A$.  What is
equilibrium in this model? Intuitively, $n_{eq}=N/2$ as it corresponds
to the state which can be realized with the largest number of
microscopic configurations. Like in the free expansion, at equilibrium
the gas fills equally (on average) both halves of the container.  The
simplicity of the model allows us to monitor the evolution of an
ensemble of initial conditions starting from state $n_0$ by
analytically computing the evolution of $\langle n_t \rangle$ and
$\sigma_t^2=\langle n^2_t \rangle-\langle n_t \rangle^2$, introducing
$\Delta_0=n_0-N/2$:
\begin{eqnarray}
\langle n_{t}\rangle\!\! &=&\!\!\frac{N}{2}+\Bigl( 1 - \frac{2}{N} \Bigr)^t \Delta_0 \label{eq:nt}\\
\sigma_t^2\!\! &=&\!\! \frac{N}{4} \!+\!\left(1\!-\!\frac{4}{N}\right)^t
\!\!\left(\Delta_0^2\!-\!\frac{N}{4}\right)\! -\!\left(1\!-\!\frac{2}{N}\right)^{2t}\!\Delta_0^2 \,.\label{eq:st}
\end{eqnarray}

Essentially, Eq.~(\ref{eq:nt}) tells us that $\langle n_t \rangle \to
n_{eq}=N/2$ exponentially fast with a characteristic time
$\tau_c=-[\ln(1- \frac{2}{N})]^{-1} \simeq N/2$, while
Eq.~(\ref{eq:st}) implies that also the standard deviation $\sigma_t$
goes to its equilibrium value $\sigma^{eq}=\sqrt{N}/2$ with a
characteristic time $O(N/2)$.  This is fine at the level of the
(ensemble) average behavior, what can we tell for the single
trajectory?

It is easy to see that  the single trajectory is also
``typical'' in the sense (\ref{eq:typ1}), i.e. 
it should basically behave as the average
trajectory, at least, if $N$ is large enough.  Consider a
far-from-equilibrium initial condition, $n_0 \simeq N$: it is easy to
prove that, if $N \gg 1$, until a time $O(N/2)$, i.e. as long as $n_t$
remains far from $n_{eq}$ each single realization of $n_t$ stays
``close'' to its average.  Indeed, Chebyshev inequality sets the bound
\begin{equation}
\mathrm{Prob}\left( {|n_t - \langle n_t \rangle| \over \langle n_t \rangle} >\epsilon \right)\le {\sigma_t^2 \over \langle n_t \rangle^2 \epsilon^2}\,\,.
\label{eq:typ}
\end{equation}
for the probability that $n_t$ deviates from its mean more than a
small percentage $\epsilon$.  From Eq.\eqref{eq:st} and
Eq.\eqref{eq:nt}, we obtain the bound $\sigma_t^2/\langle n_t \rangle
=\mathcal{O}(1/N)$. Then, back to Eq.~(\ref{eq:typ}) we have that for
every $\epsilon$ at will, there exists an $N_\epsilon$ such that, with
probability $\approx 1$, each $n_t$ stays close its average if $n_t -
n_{eq} \gg \sqrt{N_\epsilon}$ (i.e. at time $t$ the system is still
far from equilibrium). The above result means that we will observe an
irreversible tendency to reach the equilibrium value in any single
trajectory.  Conversely, if $n_0 \sim n_{eq}$,
i.e. $|\Delta_0|=|n_0-n_{eq}| \sim \sigma^{eq}$ we cannot distinguish
the initial condition from a spontaneous fluctuation from equilibrium.

In the Ehrenfest model, it is possible to show that for $N\gg 1$ and
far enough from equilibrium (i.e. $|n_0-N/2|\gg O(\sqrt{N})$), both
the Zermelo and Loschmidt paradoxes (suitably reinterpreted in the
context of this Markov chain model) are physically irrelevant, see
Ref. \cite{KAC57} for a detailed discussion.

The Ehrenfest urn-model is a useful example 
to illustrate some basic aspects of Boltzmann's viewpoint, 
even though the stochastic nature of the model
might seem too far from the ``mechanical context'' where
irreversibility is traditionally discussed. 
Nevertheless, this model maintains some similarities with
deterministic Hamiltonian systems.  For instance it is easy to show
that it satisfies the detailed balance property
$P(n_t=i;n_{t+1}=j)=P(n_t=j;n_{t+1}=i)$, that is the stochastic
equivalent of microscopic reversibility \cite{risken1984fokker}.  In
the following, we present numerical examples of Hamiltonian systems
showing the scenario here discussed remains basically unchanged also
in the deterministic world.

\section{Irreversibility in large deterministic Hamiltonian systems
\label{sec:3}}
In this section, we study two examples of many
particle Hamiltonian systems in which the volume available to the
particles is constrained along a direction by a moving wall (a
piston). The position of the piston is a macroscopic observable,
corresponding to the volume occupied by the system at a certain time,
and therefore can exhibit an irreversible behavior when initialized
in a non-equilibrium state. We will consider both
interacting and non-interacting particles. However, we emphasize that
even when the gas particles do not interact directly, 
they do it indirectly via the collisions with the moving wall (piston).

\subsection{A mechanical model of thermometer
\label{sec:termometro}}

We start from the following mechanical model: a pipe, containing $N$
particles of mass $m$, is horizontally confined, on the left, by a
fixed wall and, on the right, by a wall free to move without friction
(the piston), of mass $M$, whose position changes due to collisions
with the gas particles and under the action of a
constant force $F$.  We consider two actualizations of the system with
and without direct interaction among particles. 
As discussed in the following, the latter system is chaotic while the former 
is not, therefore their comparison provides a test on the role of chaos in 
macroscopic irreversibility.

In the non-interacting gas case, the Hamiltonian reads
\begin{equation}
\label{eq:HP_NI}
\mathcal{H}_0=\sum_{i=1}^{N}\frac{\bm p_i^2}{2m}+ \frac{P^2}{2M} + FX\,,
\end{equation}
plus terms accounting for the interactions with the walls against
which the particles collide elastically. Particle momenta are denoted
with $\bm p_i$ while $X$ and $P$ are the piston position and momentum,
respectively.

The equilibrium statistical properties of the 
system can be easily computed using the microcanonical
ensemble~\citep{FALCIONI11,CERINO14}.  At equilibrium, the gas
particles are uniformly distributed within the available volume, in
particular the horizontal coordinate $x_i$ is uniform in $[0:X]$, with
velocities following the Maxwell-Boltzmann distribution at the
equilibrium temperature, $T_{eq}$.  We can easily compute the
equilibrium values, 
\begin{equation}
T_{eq}=\frac{2NDT_0+2FX_0}{N(D+2)+3}\,,\; X_{eq}=(N+1) \frac{T_{eq}}{F}\,,\;
\sigma^{eq}_X=\alpha(D)\sqrt{N}\frac{T_{eq}}{F} + o(\sqrt{N})
\label{eq:eqtermomento}
\end{equation}
where $\alpha(D)$ is a constant depending on the system
dimensionality $D$: $\alpha(D=1)=1/\sqrt{3}$,  
$\alpha(D=2)=1/\sqrt{2}$ and from now on we work in units such that $k_B=1$.
Eqs.~(\ref{eq:eqtermomento}) show that the piston position provides a 
measure of the temperature, once $F$ and $N$ are given.
We notice that the average becomes more and more sharp, 
$\sigma^{eq}_X/X_{eq}=O(N^{-1/2})$, as $N$
increases. It is worth emphasizing that, in the absence of
interactions, the horizontal axis is the only relevant direction.  For
this reason numerically we have studied it in one dimension.

We conclude the presentation of the non-interacting gas model by
emphasizing that the whole system is not chaotic, i.e. it has
vanishing Lyapunov exponents. The dynamics of the non-interacting gas
plus piston can be mapped into that of billiard whose boundary is a
polyhedron, and thus with zero curvature.  It is a known fact that for
billiards in with zero-curvature boundaries (and thus, for our
mechanical model) all Lyapunov exponents do vanish, though the system
can still be ergodic \cite{lebo}.

In the interacting gas case, we consider a two-dimensional 
pipe, of cross-section $L$, with $\bm q_i=(x_i,\,y_i)$ and $\bm
p_i=(p^x_i,\,p^y_i)$.  The Hamiltonian is obtained by adding to Eq.~(\ref{eq:HP_NI})
the interaction potential, so that
\begin{eqnarray}
\mathcal{H} = 
\mathcal{H}_0+\sum_{i<i'}  U(|\bm q_i-\bm q_{i'}|) + 
 U_{w}(\bm q_1,\dots,\bm q_n, X)\,.\label{eq:HP_I}
\end{eqnarray}
We consider repulsion between particle pairs, $U(r_{ij})=U_0
/r_{ij}^{12}$, and with the four walls, $U_{w}(R)=U_0/R^{12}$, where
$R$ denotes the particle-wall distance. The right wall is the
frictionless piston. Previous numerical
investigations \cite{CERINO14} have shown that at low densities the
system behaves like a two-dimensional  ideal gas.
From a quantitative point of view, there will be corrections (whose calculation is not of interest here) with respect to the equilibrium values (\ref{eq:eqtermomento}) (for $D=2$)  due to the interaction energy.
Interestingly for our discussion here, the major qualitative difference with
respect to the non-interacting gas is a dynamical one: due to the
non-linear interactions among the particles, now the system is
chaotic, as demonstrated in Fig.~\ref{fig:lyap}.
\begin{figure}[!b]
\centering
\includegraphics[clip=true,width=8.6cm]{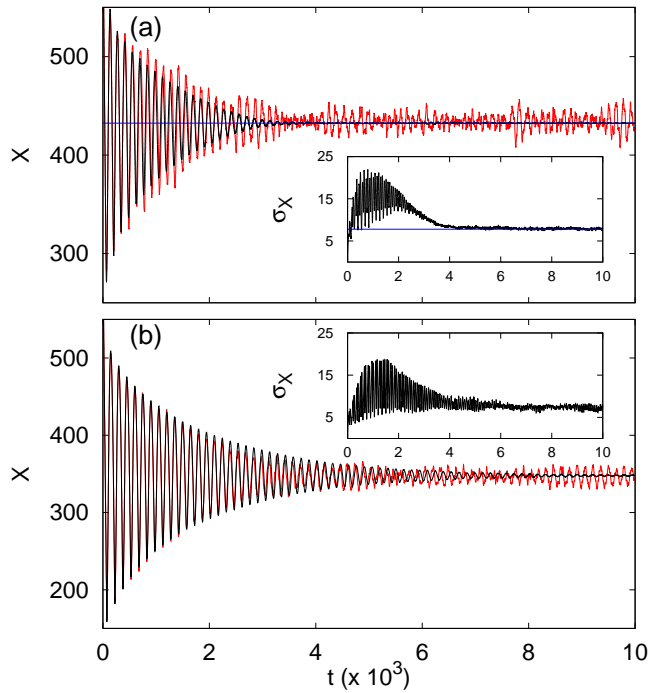}
\caption{(color online) Irreversibility in the thermometer model: 
piston position in time for the non-interacting (a) and interacting
(b) gas.  Black and red curves denote ensemble averaged and
single-realization trajectory, respectively. Blue
horizontal lines denote the equilibrium values in the one-dimensional
non-interacting case.  Insets: standard deviation, $\sigma_X(t)$.
Simulation parameters: $N=1024$, $m=1$, $M=50$, $F=150$, $T_0=10$,
$X_0=600$.  In the interacting case the pipe cross-section is $L=30$
and the interaction intensity $U_0=1$.  Averages are over $2000$ (a)
and $150$ (b) realizations. \label{fig:1}}
\end{figure}

\begin{figure}[!hbtp]
\centering
\includegraphics[clip=true,width=8.6cm]{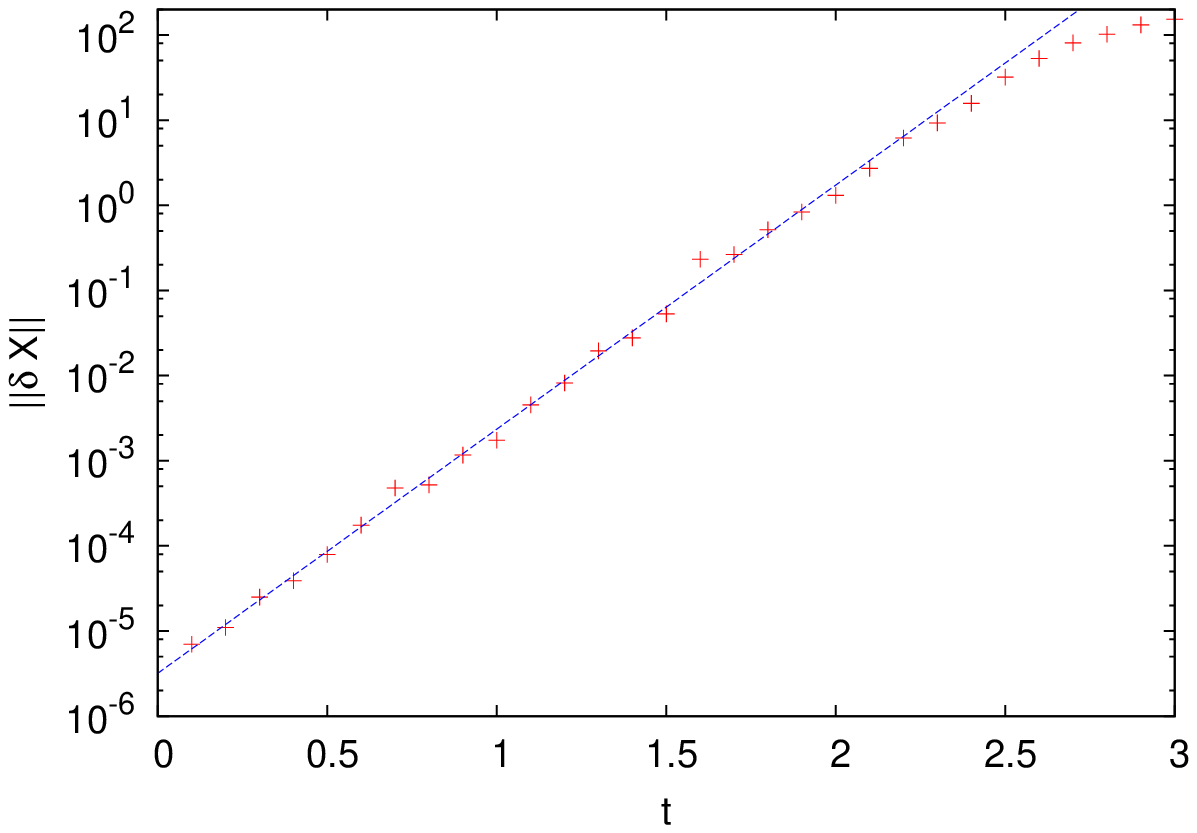}
\caption{(color online) Sensitive dependence 
on initial conditions in the interacting gas model. Typical 
evolution of the distance between two trajectories of
the system starting from two close initial conditions $\bm X(0)$ and
$\bm X'(0)$, with $ ||\delta {\bm
X}(0)||=\left(\sum_{i=0}^{2N} (X_i'(0)-X_i(0))^2\right)^{1/2}=10^{-6}$.
The distance $ ||\delta {\bm
X}(t)||$ increases exponentially with time with rate given by 
the maximal Lyapunov exponent, $\lambda_1\approx 6$. All the parameters are
the same of Fig.\ref{fig:1}\label{fig:lyap}.} 
\end{figure}
We now discuss irreversibility by following the time evolution of
the piston position in the interacting and non-interacting case.  At
time $t=0$, we fix the position of the piston $X(0)=X_0$, its velocity
$V(0)=0$, and set the initial microscopic state as an equilibrium
configuration of the gas in the volume imposed by the piston position
at a given temperature $T_0$. In practice, we take the gas particles
uniformly distributed in $[0:X_0]$ (in the two-dimensional case, in
$[0:X_0]\times[0:L]$) with a Maxwell-Boltzmann distribution of
velocities at temperature $T_0$.

We run molecular dynamics simulation by using event-driven schemes in
the non-interacting gas and Verlet algorithm with
time step $\Delta t=10^{-3}$ in the interacting one (see caption of
Fig.~\ref{fig:1} for specific parameters). As
expected, numerical simulations show that when the initial state is
sufficiently far from from equilibrium, meaning that
$|X_0-X_{eq}| \gg \sigma_X^{eq}$, its evolution $X(t)$ exhibits an
irreversible behavior.

Figure~\protect\ref{fig:1}a reports a single trajectory, $X(t)$, and
the behavior of the ensemble average, $\langle X(t)\rangle$, obtained
by repeating the simulation from the same macroscopic initial
condition (the same $X_0$ and $T_0$) but different microscopic
initializations of the gas particles. We fixed $|X_0-X_{eq}| \approx
10\, \sigma_X^{eq}$. In analogy with the Ehrenfest model, we observe
that the average trajectory is also typical: far from equilibrium,
fluctuations are small compared to the ensemble average value.  In
other words, for almost every initial configuration of the system
compatible with the macroscopic state, the time evolution of the
piston position is practically identical to the average
one. The standard deviation of the position, $\sigma_X(t)$,
 as shown in the inset of Fig.~\ref{fig:1}a, evolves from the initial value  $0$ (by construction) and reaches the equilibrium
value at long time, similarly to the Ehrenfest model but with a richer
and more complex phenomenology.  In particular, we notice the
non-monotonic behavior of $\sigma_X(t)$ in the short-time oscillatory
phase.  Similar behaviors are not uncommon for systems starting from
an unstable state~\cite{TOMBESI81}. However, and interestingly for
our discussion, $\sigma_X(t)$ remains small with respect to the
average value. We can thus observe macroscopic irreversibility in a
single trajectory of macroscopic system initialized in a
non-equilibrium initial state.

The interacting particle system
(\ref{eq:HP_I}) qualitatively displays the same
behavior (Fig.~\ref{fig:1}b) supporting the statement
that (microscopic) chaos does not add any
new relevant feature to macroscopic irreversibility.
\begin{figure}[!thb]
\centering
\includegraphics[clip=true,width=8.6cm]{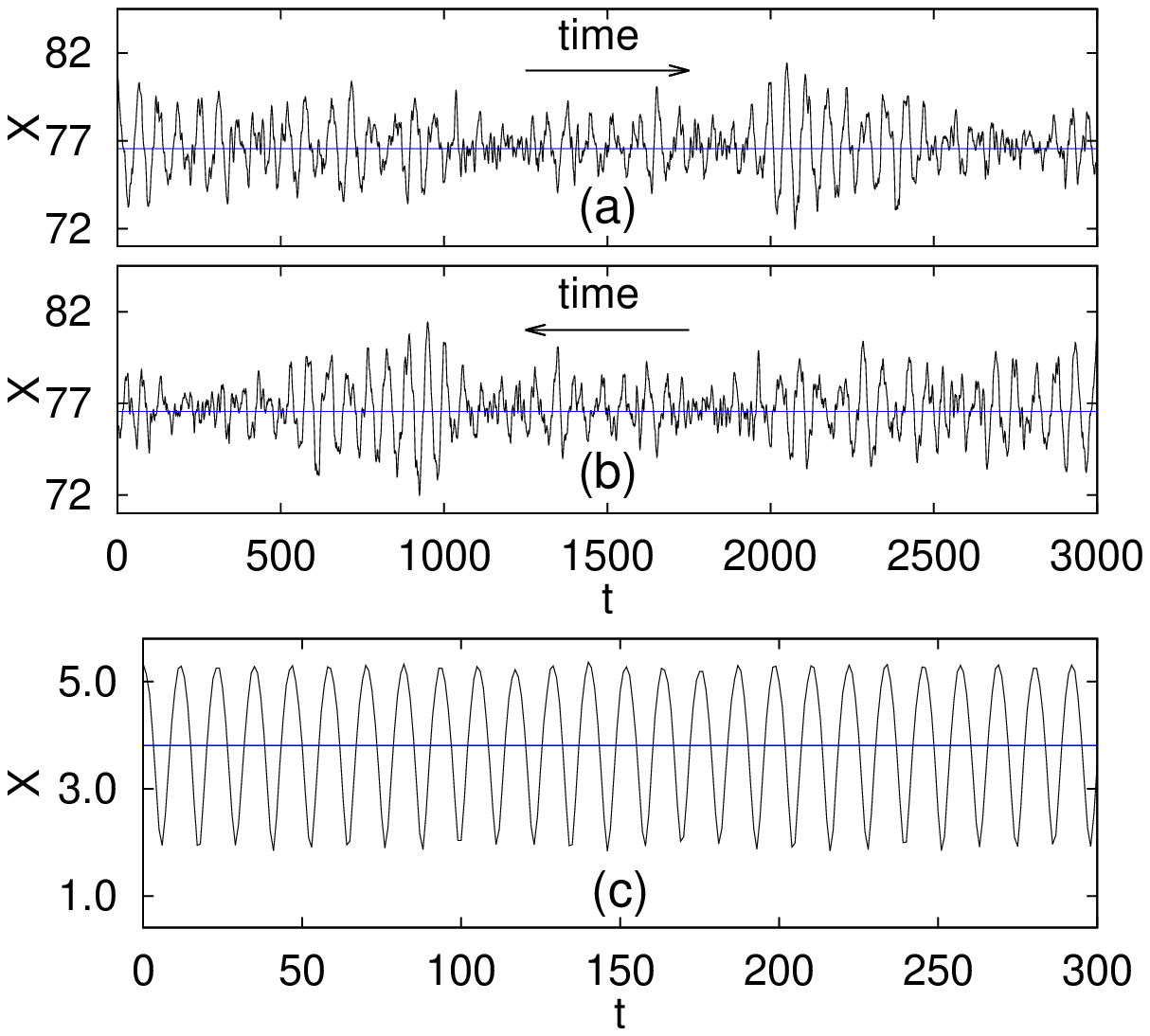}
\caption{(color online) Evolution from close-to equilibrium initial conditions, 
(a) and (b), or for small systems (c). Piston position $X$ vs time for
the non-interacting gas system: (a) with $N=1024$, $T_0 =10$ and
$X_0=X_{eq}+3\,\sigma_X^{eq}$; (b) the time reversed trajectory of
(a), as marked by the arrows; (c) with $N=4$, $T_0 = 10$, $M=40$,
$F=15$ and $X_0=1.4\cdot T_{eq}$.  Horizontal (blue)
lines denote the microcanonical ensemble average position of the
piston $X_{eq}$. Other parameters are as in
Fig.~\ref{fig:1}.\label{fig:2}}
\end{figure}

Figure~\ref{fig:2}a displays the typical evolution from a (close-to)
equilibrium initial condition,
i.e. $|X_0-X_{eq}| \approx \sigma_X^{eq}$, in
the non-interacting gas system.  As
one can see, irreversibility does not show up: the time reversed
trajectory is basically indistinguishable from the forward trajectory (compare Figs.~\ref{fig:2}a
and b).  Irreversibility cannot be observed also when the system is
small, i.e. the number of degrees of freedom ($N$) is small. In the
last case no notion of typicality can be defined: it is even
meaningless to speak of far-from-equilibrium initial conditions, as
fluctuations are of the same magnitude of mean
values.
Though the evolution is statistically
stationary, we cannot define a (thermodynamic) equilibrium state when
$N$ is small. Therefore, Fig.~\ref{fig:2} demonstrates
the importance of having a large number of degrees of freedom and of
starting from a very non-typical initial conditions for observing
macroscopic irreversibility.

Summarizing, when an experiment
 is conducted, in each\footnote{More precisely almost
 all.} single realization, the evolution of a macroscopic observable
 is close to the ensemble average and, in addition, it exhibits
 irreversibility, irrespectively of the presence of chaos in the
 system provided that the system is truly macroscopic ($N\gg 1$) and
 the initial condition is far (enough) from equilibrium. We remark
 that (microscopic) chaos is irrelevant also for
 dynamical transport properties close to
 equilibrium \cite{cecconi2007transport}.

\subsection{The adiabatic piston}
\begin{figure*}[!thb]
\centering
\includegraphics[clip=true,width=1\textwidth]{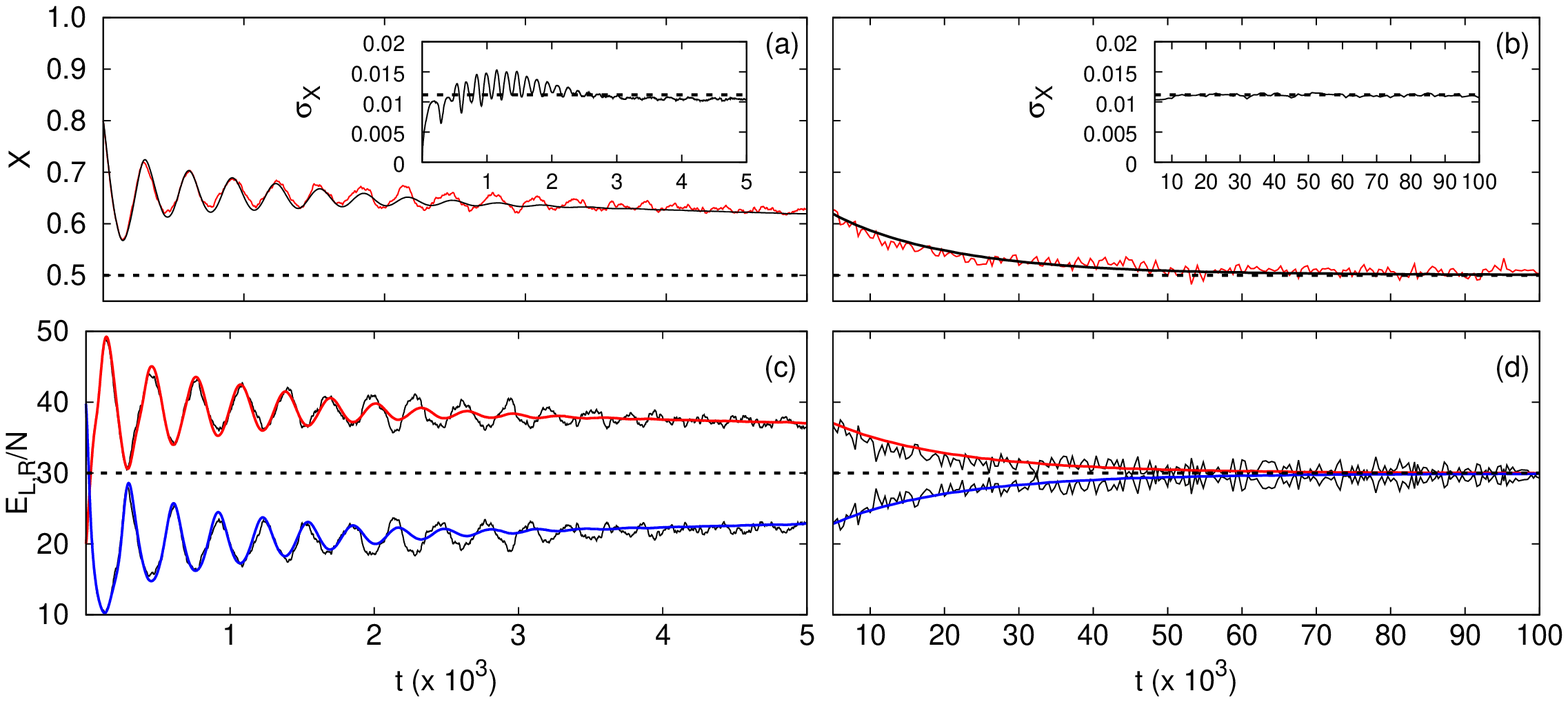}
\vspace{-0.3truecm}
\caption{(color online) Irreversibility in the adiabatic piston: (a,b)  
piston position in a single realization (red) and ensemble average
(black).  Inset: evolution of the standard deviation $\sigma_X$. (c,d)
Ensemble average of the left $E_L(t)/N$ (blue) and right $E_R(t)/N$ (red)
kinetic energy per particle, and in a single realization (black).
Horizontal (dashed) lines denote equilibrium values. The splitting in
two panels (for short times (a,c) and longer times (b,d)) is just for
an easier identification of the two regimes discussed in the text.
The simulation parameters are $N=10^3$, $m=1$, $M=100$,
$\mathcal{L}=2N$; the initial state is defined by $X_0=0.8$ with
$T_L(0)=40$ and $T_R(0)=80$, averages are on $2000$ realizations.
\label{fig:3}}
\end{figure*}

We now consider the so-called \textit{adiabatic piston} -- a classical
problem in non-equilibrium
thermodynamics~\cite{Feynman65,Lieb99,lebo,Gruber06} (see also
Ref.~\cite{Crosignani96} for a pedagogical introduction). In this
interesting example the approach to equilibrium from a non-equilibrium
state is characterized by a more complex
phenomenology than that of the previous example.

In a nutshell the system is as follows.  A thermo-mechanically
isolated cylinder of length $\mathcal{L}$ is partitioned into two
compartments by an adiabatic, freely-moving wall (the piston) of mass
$M$.  Each compartment contains a gas composed of $N$
non-interacting particles of mass $m$, elastically colliding with the
walls.  Thanks to the absence of direct interaction,
we can restrict our analysis to one dimension, along the
horizontal direction.  The system is initialized with the piston kept
fixed by a clamp at a given position, $X_0\mathcal{L}$;
the non-interacting gases in the left/right (L/R)
compartments are both in equilibrium, meaning that they are uniformly
distributed in the compartments with volumes $V_{L}(0)=X_0\mathcal{L}$
and $V_{R}(0)=\mathcal{L}(1-X_0)$, and velocities distributed with
the Maxwell-Boltzmann distribution at different temperatures
$T_{L,R}(0)$; the pressures are fixed by
the non-interacting-gas state equation $P_{L,R}(0)
V_{L,R}(0)=N T_{L,R}(0)$. Being the piston adiabatic, until the clamp
is present, the two subsystems are in equilibrium even if $T_L(0) \neq
T_R(0)$.  At $t\!=\!0$, the clamp is removed and the piston is free to
move without friction under collisions with gas particles. The
non-trivial question is to predict the final position of the piston
and values of thermodynamic quantities.

A careful treatment~\cite{curzon}, within the framework of equilibrium
thermodynamics, shows that the system should reach mechanical
equilibrium $P_L = P_R$.  However, the final position of the piston
and gas temperatures remain undetermined. The prediction of the final
equilibrium state needs to understand the
non-equilibrium process, occurring after the clamp
removal. Feynman~\cite{Feynman65} argued that the system first
converges toward a state of mechanical equilibrium with $P_R\approx
P_L$ (but for small fluctuations), consistently with the equilibrium
thermodynamic prediction. Then, pressure fluctuations,
which are asymmetric because of $T_L\neq T_R$, slowly drive the system
toward thermal equilibrium $T_R=T_L$. The final position of the piston
is $X_{eq}=1/2$ with standard deviation
$\sigma_{X}^{eq}=1/(\sqrt{8N})$ \cite{DELRE11}. The equilibrium
temperature, $T^{eq}_L=T^{eq}_R=(T_L(0)+T_R(0))/2 +O(1/N)$, can be
directly derived from the conservation of energy fixed by the initial
value $E=N((T_L(0)+T_R(0))/2$.  Despite many
attempts~\cite{morrissgruber03,Mansour,lebo,CPPV07} to derive Feynman
predictions within a consistent analytic framework is a not yet solved
problem even for non-interacting gases.

Here, our interest is to show that the scenario for macroscopic
irreversibility so far discussed well applies also to this more
complex irreversible process, characterized by the two regimes
identified by Feynman.

In Figure~\ref{fig:3}, we show the irreversible macroscopic evolution
of the system by monitoring the piston position
and the kinetic energy per particle in each
compartment $E_{L,R}(t)/N$ that, when the gases in each chamber are in
equilibrium, are nothing but half the temperature values. 
Analogously to the previous section, we show both the evolution averaged 
over many realizations with the same initial macroscopic state and a single
realization.  Panels (a,c) refer to the first stage of the evolution
ending with the equilibration of pressures; panels (b,d) pertain to
the second stage in which, while $P_L \approx P_R$, asymmetric
pressure fluctuations drive the system towards the final equilibrium
state.  The insets show the time evolution of the standard deviation
of the piston position $\sigma_X(t)$ which behaves similarly to the
thermometer model.  As clear from the figure in both regimes any
single trajectory closely traces the average one, a manifestation of
typicality as previously discussed and a further demonstration of the
validity of Boltzmann's scenario for irreversibility, also in this
non-trivial example.

\section{Spreading of an ``ink'' drop\label{sec:4}}
When an ink drop falls into the water, we observe its irreversible
spreading and mixing with the fluid. A typical way to describe the
phenomenon is in terms of the diffusion equation. The idea underlying
such approach is to mimic the collisions of an ink molecule against
water molecules by a stochastic force, renouncing to a deterministic
description. Another possibility, within the deterministic framework,
is to use molecular dynamics, but this can be very heavy from a
computational point of view.

Here, we introduce an idealized simple model which can be used to
study such a phenomenon from a conceptual point of view. We study a
discrete-time high dimensional symplectic map (akin to a high
dimensional Hamiltonian system) involving $2N$ degrees of freedom, and
$2$ auxiliary variables. We consider a special case of the system
proposed in Ref.~ \cite{Boffetta03}, in particular
\begin{equation}
\label{mappette}
\left\{
\begin{array}{lll}
y_i(t+1) &=& y_i(t) + \epsilon\cos[x_i(t)- \theta(t)]\\ x_i(t+1) &=&
x_i(t) + y_i(t+1) \\ J(t+1) &=& J(t)
- \epsilon \sum_{j=1}^N\cos[x_j(t)- \theta(t)],\\
\theta(t+1) &=& \theta(t) + J(t+1) \,.
\end{array}
\right. 
\end{equation}
Each pair $(x_i,y_i)$ identifies a ``particle'' ($i=1,\ldots,
N$),\footnote{Notice that $x_i$ and $y_i$ can be
interpreted as the position and momentum of the $i$-particle,
respectively.} and periodic boundary conditions on the
two-dimensional torus $\mathbb{T}_2 = [0,2\pi]\times[0,2\pi]$ are
assumed. For $\epsilon=0$, the particles do not
interact, while when $\epsilon>0$ (in our numerical
examples we use $\epsilon=1$) particles interact (the ``collisions''
of water molecules) via a mean-field-like interaction, mediated by the
variables $\theta$ and $J$.  We emphasize that $\theta$ and $J$ do not
have a precise physical meaning, they represent a simple mathematical
expedient to introduce the interaction among particles in a symplectic
manner. Moreover, the mean field character of the interaction is
immaterial here and it simply allows fast numerical
computation.  In the presence of interactions the system exhibits
complex evolutions, as realistic gases or liquids in molecular
dynamics systems.  System (\ref{mappette}) can be shown to be 
time-reversible, see Ref.~\cite{quispel92} for a detailed discussion on 
time reversal symmetries of discrete-time dynamical systems).
\begin{figure}[!t]
\centering
\includegraphics[clip=true,width=8cm]{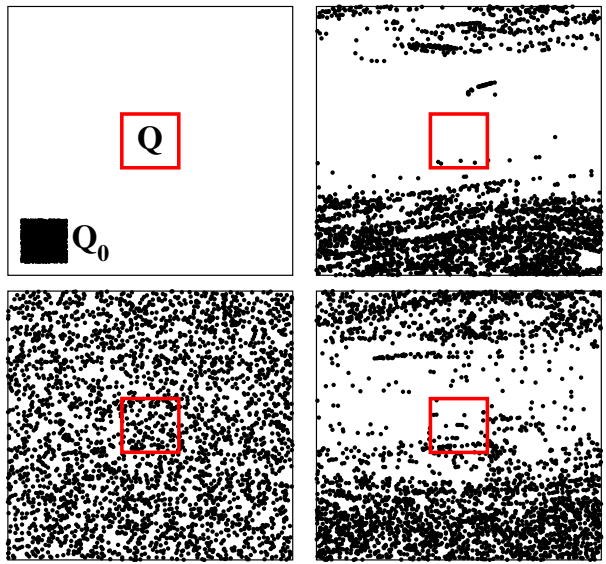}
\caption{(color online) Irreversible spreading of an ink drop
 of $N_I=3.2\cdot 10^3$ particles on the Torus $\mathbb{T}_2$ at
$t=0,4\cdot 10^3,2.9\cdot 10^4,2.33\cdot 10^5$ (in clockwise order
from top left). The $N_I$ ink particles start uniformly distributed in
$Q_0 \equiv [0.3:1.3]\times[0.3:1.3]$, while the $N_W=10^7$ solvent
ones have been thermalized in a previous time integration.  The
instantaneous occupation $n(t)$ is monitored in the (red) box $Q$
centered in $(\pi,\pi)$ with side $\pi/5$.
\label{fig:4a}}
\end{figure}
  We used a system with interacting particles to avoid confusion
between the genuine thermodynamic irreversibility and the mixing
property, Eq. (\ref{eq:mixing}). As already stressed
in Sec.~\ref{sec:2a}, since our system is composed of $N$ interacting
elements it should be clear that we are dealing with a single large
system and not with a collection of different initial conditions as if
the particles were non-interacting and evolving according to a generic
mixing map of the torus. In this respect, we emphasize that the details
of the interaction among the particles are not particularly important
provided some form of interaction is present.

After several iterations, the system \eqref{mappette} reaches an
``equilibrium'' dynamical state characterized by a uniform
distribution of particles on $\mathbb{T}_2$.  To mimic the spreading
of a cloud of ``ink'', we split the $N$ particles into $N_W$ particles
of solvent (water) and $N_I$ particles of solute (ink), with $N = N_W
+ N_I$ and $N_I\ll N_W$.  Then, we prepare the initial condition of
the system with the $N_W$ particles at equilibrium (e.g. after a long
integration with $N_W$ particles only), and the solute particles
uniformly distributed in a small region $Q_0$ of $\mathbb{T}_2$ (top
left panel in Fig.~\ref{fig:4a}). During the evolution, to measure the
degree of mixing, we monitor the number of ink particles, $n(t)$,
which at time $t$ reside in a given set $Q \subseteq \mathbb{T}_2$
(the red box $Q$ in Fig.~\ref{fig:4a}). At equilibrium, when ink is
well mixed, the $N_I$ particles will also distribute uniformly over
$\mathbb{T}_2$, and thus $n(t)$ will fluctuate around $n_{eq}=
N_I \mathcal{A}(Q)/\mathcal{A}(\mathbb{T}_2)$, where $\mathcal{A}(Q)$
is the area of the subset $Q$.

\begin{figure}[!t]
\centering
\includegraphics[clip=true,width=8cm]{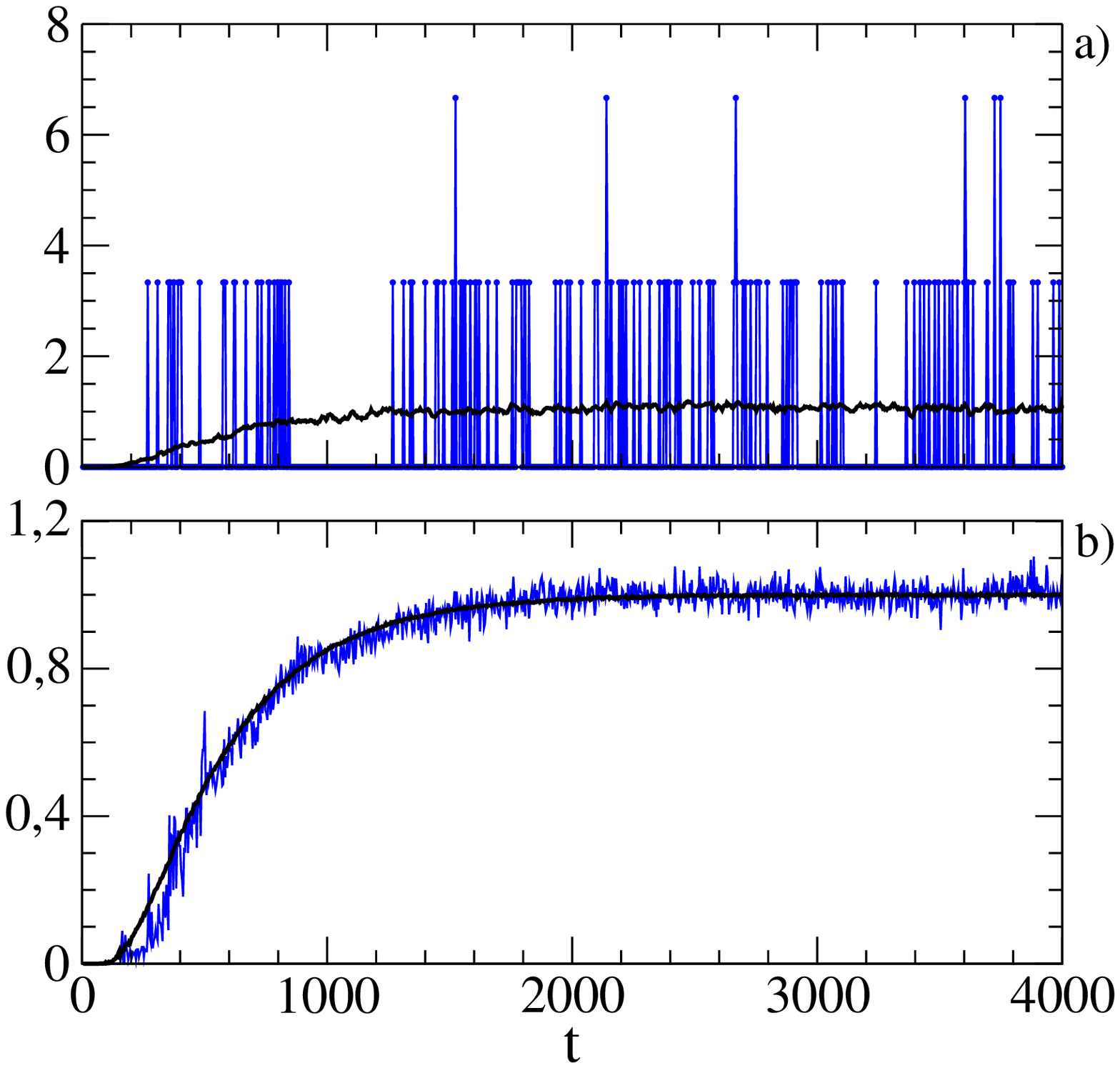}
\caption{Instantaneous occupation $n(t)/n_{eq}$ of the set $Q$ (blue,
fluctuating curve) and its average $\langle n(t) \rangle/n_{eq}$
(black, smooth curve) over 500 independent initial conditions starting
from $Q_0$: (a) $n_{eq}=0.3$ (drop with very few particles, $N_I=8$
and $N_W=2500$) and (b) $n_{eq} = 10^3$ (drop with many particles
$N_I=2.5\times 10^4$ and $N_W=10^6$).\label{fig:4b}}
\end{figure}

It is instructive to compare (see Fig.~\ref{fig:4b}) the behavior of
$n(t)$ for a single trajectory with the average $\langle
n(t) \rangle$, computed over an ensemble of many independent releases
of the ink drop, with the water in
different (microscopic) initial
conditions arbitrarily chosen in the equilibrium
state.  Moreover, we study the difference between the case $N_I \sim
O(1)$ (Fig.~\ref{fig:4b}a) and $N_I \gg 1$ with $N_I\ll N_W$
(Fig.~\ref{fig:4b}b). It is important to realize that
while the latter case ($N_I\gg 1$) the ink drop can be considered a
macroscopic object, in the former ($N_I\sim O(1)$) it cannot.  In
both cases, we observe that $\langle n(t)\rangle/n_{eq}$ increases
monotonically with $t$, asymptotically approaching $1$. However, a
dramatic difference emerges if we look at the single realization.  For
a (macroscopically well defined) drop with $N_I \gg 1$, the single
trajectory closely follows the average one (Fig.~\ref{fig:4b}b), and
we can define an irreversible behavior for the
individual drop. Conversely, when $N_I \sim O(1)$, the single
trajectory is indistinguishable from its time reverse one
(Fig.~\ref{fig:4b}a) and strongly differs from the average one. The
latter apparently shows a form of irreversibility, but it is thus a
mere artifact of the average over the initial distribution and the
special initial condition.  
We stress that, the lack of irreversibility in this case is due to 
the fact that, being $N_I$ small, $n(t)$ cannot be considered a 
macroscopic observable even if the system water plus drop is large ($N\gg1$),
as $n(t)$ depends only on the few ``molecules'' of ink.

\section{Final Remarks}\label{sec:5}

In this work,
 with the help of numerical simulations of simple, yet non-trivial,
Hamiltonian models, we revisited some of the basic aspects of
Boltzmann's interpretation of irreversibility. It is worth concluding by
listing some of the key elements underlined by our simple investigation.
\begin{enumerate}
\item Irreversibility is observed and must be defined in a single
macroscopic body.  This implies that averaging over
all the possible initial conditions is unnecessary both at a practical
and conceptual level, as perfectly obvious to experimentalists.

\item  Crucial to observe irreversibility is the choice of 
the initial condition, which has to be very ``unlikely'', that is sufficiently 
far-from equilibrium. Indeed, even in a large-$N$ system,
irreversibility does not show up in a trajectory starting from initial
conditions chosen close-to-equilibrium (see Fig.~\ref{fig:2}a and
b).

\item  Irreversibility is a property of macroscopic bodies, 
i.e. of system with a large number of components $N\gg 1$.  Indeed,
the large $N$ condition of a system grants that it develops a
``typical'' behavior, meaning that the features of a single system are
close to their averages. 
\item The
presence, or absence, of chaos is not relevant. Chaos plays a role in
mixing, which is surely a form of ``irreversibility'', but which has
nothing to do with the second law.
\end{enumerate}
All the irreversible behaviors in the approach to equilibrium that we
observed in the examined examples clearly confirm the above conceptual
framework whenever the system is composed of a large number of
particles and the observables are macroscopic, i.e. depend upon a
large number of degrees of freedom.  Conversely, when either the number of
particles is small or the observed quantity depends on few
degrees of freedom,  we are unable to identify a clear
trend towards equilibrium and we cannot determine the time arrow. 
 
\section*{Acknowledgements}
We thank M. Falcioni for a critical reading of the manuscript and
useful remarks.

\bibliographystyle{elsarticle-num}
\bibliography{bibli}

\end{document}